\documentclass[preprint,showpacs]{revtex4}
\usepackage{amssymb}
\usepackage{amsmath}
\usepackage{graphicx}

\def\be{\begin{eqnarray} &&}

\def\ee{\end{eqnarray}}

\def\bew{\begin{widetext}}
\def\ew{\end{widetext}}

\begin{document}

\title{Light scalar tetraquarks from a holographic perspective}
\author{Hilmar Forkel$^{1}$}
\affiliation{$^{1}$Institut f\"{u}r Physik, Humboldt-Universit\"{a}t zu Berlin, D-12489
Berlin, Germany }

\begin{abstract}
We discuss how a dominant tetraquark component of the lightest scalar mesons
may emerge in AdS/QCD gravity duals. In particular, we show that the
exceptionally strong binding required to render the tetraquark ground state
lighter than the lowest-lying scalar quark-antiquark nonet can be
holographically encoded into bulk-mass corrections for the tetraquark's dual
mode. The latter are argued to originate from the anomalous dimension of the
corresponding four-quark interpolator. To provide a concrete example, we
implement this mechanism into the dilaton soft-wall dual for holographic
QCD. Preventing the lowest-lying dual mode from collapsing into the AdS
boundary then establishes a rather generic lower bound on the tetraquark
mass (which may be overcome in the presence of additional background
fields). We further demonstrate that the higher tetraquark excitations can
become heavier than their quark-antiquark counterparts and are thus likely
to dissolve into the multiparticle continuum.
\end{abstract}

\pacs{11.25.Tq, 11.25.Wx, 14.40.Cs, 12.40.Yx}
\maketitle


\section{Introduction}

The nature of the light scalar mesons remains a key challenge for our
understanding of how QCD generates bound states. Exceptionally strong and
diverse mixing patterns between the differently flavored quarkonium
components and with scalar glueballs are among the intricacies which kept
this sector exciting and controversial for more than four decades.
Theoretical approaches to the subject include a long tradition of model
building, methods based on dispersion relations and effective field
theories, QCD sum-rule analyses and more recently devoted and unquenched
lattice simulations. The corresponding literature may be\ traced at least
partly from the reviews \cite{clo02,ams04,kle07,ams08,latt}.

In recent years, evidence has accumulated in favor of Jaffe's classic
suggestion \cite{jaf77} that the lightest scalar meson nonet with masses
below 1 GeV (and potentially a first recurrence beyond 1 GeV \cite{mai07})
may contain a dominant (cryptoexotic) tetraquark component \cite%
{mai04,bri05,jaf05,tho08,ams08,latt}. This at present arguably most
convincing interpretation demands an explanation for the exceptionally
strong binding between the four \textquotedblleft valence\textquotedblright\
quarks, however, which is able to reduce the lightest tetraquark mass
significantly below that of the scalar quark-antiquark ground state. With
the advent of holographic strong-coupling techniques based on the
gauge/gravity correspondence \cite{revs1} and the construction of a first
generation of approximate holographic QCD (or AdS/QCD) duals \cite{revs2} it
seems pertinent to explore whether those can provide new insights into
existence and structure of strongly bound tetraquarks. In the present note
we will take a first step in this direction.

Several holographic studies of the light spin-0 meson \cite%
{veg08,col08,dep09} (and glueball \cite{bugb}) sectors in the AdS/QCD
framework have recently appeared. Those were based exclusively on
quark--antiquark interpolators, however, which presuppose a $\bar{q}q$%
-dominated meson structure. Moreover, the first two of them exclude the
lightest scalars ($f_{0}\left( 600\right) $ and $K_{0}^{\ast }\left(
800\right) $) from their analysis. In any case, it is rather obvious (and
will be discussed in more detail below) that a straightforward extension of
this approach to $\bar{q}^{2}q^{2}$ interpolators\ would describe four-quark
states which are heavier, not lighter than the $\bar{q}q$ ground state. Our
main focus will therefore be to search for a minimal but systematic AdS/QCD
extension in which tetraquarks emerge naturally as the lightest scalar
mesons. The sought-after holographic mechanism should further be able to
render higher tetraquark excitations heavy (and thus broad) enough to avoid
generating more low-lying scalar nonets than experimentally seen.

In looking for a holographic representation of the required binding, we are
guided by the expectation that the lightest scalar tetraquarks essentially
consist of a \textquotedblleft good\textquotedblright\ (i.e. maximally
attractive) diquark and a good antidiquark \footnote{%
One should keep in mind, however, that alternative descriptions in terms of
molecular\ resonances (see e.g. Ref. \cite{mei91} and the cited reviews) are
difficult to distinguish both in QCD \cite{jaf05} and holographically.}
which form an $s$-wave bound state \cite{jaf77,mai04,jaf05}. The exceptional
lightness of the tetraquark ground state is then mainly a consequence of the
good diquark's large (and potentially instanton-induced \cite%
{sch94,sch03,cri04}) binding energy \cite{sel06}. This suggests to adapt the
approach introduced in Ref. \cite{for09}, which describes systematic mass
reductions among the light baryons due to good-diquark binding
holographically in terms of bulk-mass corrections for their dual modes, to
the problem. Below we will explore to what extent such corrections, as
induced by the anomalous dimension of the corresponding four-quark
interpolator, are able to generate light scalar tetraquarks as well.

\section{Scalar soft-wall dynamics}

As a specific dynamical basis for our study we adopt the dilaton-soft-wall
gravity dual of Ref. \cite{kar06}\ which is governed by the action%
\begin{equation}
S^{\left( \text{sw}\right) }=\frac{1}{\kappa ^{2}}\int d^{5}x\sqrt{%
\left\vert g\right\vert }e^{-\Phi }tr\left\{ \left\vert DX\right\vert
^{2}-m_{5}^{2}\left\vert X\right\vert ^{2}+\frac{1}{4g_{5}^{2}}\left(
F_{L}^{2}+F_{R}^{2}\right) \right\} .  \label{act}
\end{equation}%
The gauge-covariant derivative $D_{M}=\partial _{M}-iA_{L,M}+iA_{R,M}$
contains the left/right flavor gauge potentials $A_{L/R,M}$ with field
strengths $\left( F_{L/R}\right) _{MN}=\partial _{M}A_{L/R,N}-\partial
_{N}A_{L/R,M}-i\left[ A_{L/R,M},A_{L/R,N}\right] $. The spin-0 bulk field 
\begin{equation}
X=\left( X_{0}+\varphi ^{0}t^{0}+\varphi ^{a}t^{a}\right) e^{2i\pi ^{a}t^{a}}
\end{equation}%
consists of the background field $X_{0}$ dual to the quark condensate, a
flavor nonet $\varphi ^{A}$ (with $A\in \left\{ 0,...8\right\} $) of scalar
fluctuations (where the flavor singlet $\varphi ^{0}$ may contain a glueball
component) and the chiral pseudo-Goldstone bosons $\pi $. The U$\left(
3\right) $ flavor generators are normalized as $tr\left\{ t^{A}t^{B}\right\}
=\delta ^{AB}/2.$ The remaining, non-dynamical background fields comprise
the AdS$_{5}$ geometry 
\begin{equation}
ds^{2}=g_{MN}dx^{M}dx^{N}=\frac{R^{2}}{z^{2}}\left( \eta _{\mu \nu }dx^{\mu
}dx^{\nu }-dz^{2}\right)  \label{met}
\end{equation}%
with curvature radius $R$ (where $\eta $ is the four-dimensional Minkowski
metric and $z$ the Poincar\'{e} coordinate of the fifth dimension) and the
dilaton field%
\begin{equation}
\Phi \left( z\right) =\lambda ^{2}z^{2}
\end{equation}%
which is solely responsible for conformal-symmetry breaking in the infrared.
The mass spectrum of the scalar meson nonets is then determined by the $%
\varphi $-dependent, bilinear part 
\begin{equation}
S\left[ \varphi \right] =\frac{1}{2\kappa ^{2}}\int d^{5}x\sqrt{\left\vert
g\right\vert }e^{-\Phi }\left[ g^{MN}\partial _{M}\varphi \partial
_{N}\varphi -m_{5}^{2}\varphi ^{2}\right]  \label{bilact}
\end{equation}%
of the bulk action (\ref{act}). Since the dynamics (\ref{bilact}) is flavor
diagonal, we have dropped the index of the scalar fields. (The
\textquotedblleft inverted\textquotedblright\ flavor-breaking pattern in
tetraquark\ nonets \cite{jaf05} is thus beyond the scope of the present
study.)

Variation of the action (\ref{bilact}) generates the field equation%
\begin{equation}
\left[ \nabla _{M}\nabla ^{M}-\left( \partial _{M}\Phi \right) \partial
^{N}+m_{5}^{2}\right] \varphi =0  \label{deq1}
\end{equation}%
where $\nabla _{M}\varphi =\left( \partial _{M}+\Gamma _{MN}^{N}\right)
\varphi $ and $\Gamma _{MN}^{N}=\partial _{M}\ln \left\vert g\right\vert
^{1/2}$. In terms of the four-dimensional Fourier transform%
\begin{equation}
\varphi \left( x,z\right) =\int \frac{d^{4}q}{\left( 2\pi \right) ^{4}}%
e^{-iqx}\hat{\varphi}\left( q,z\right)
\end{equation}%
the ensuing radial equation becomes 
\begin{equation}
\left[ \partial _{z}^{2}-\left( \frac{3}{z}+2\lambda ^{2}z\right) \partial
_{z}+q^{2}-\frac{R^{2}m_{5}^{2}}{z^{2}}\right] \hat{\varphi}\left(
q,z\right) =0.  \label{radeq}
\end{equation}%
Redefining the scalar field as%
\begin{equation}
\hat{\varphi}\left( q,z\right) =\left( \frac{z}{R}\right) ^{\frac{3}{2}%
}e^{\lambda ^{2}z^{2}/2}\phi \left( q,z\right)
\end{equation}%
then eliminates the first-derivative term and turns Eq. (\ref{radeq}) into
the Sturm-Liouville problem 
\begin{equation}
\left[ -\partial _{z}^{2}+V\left( z\right) \right] \phi \left( q,z\right)
=q^{2}\phi \left( q,z\right)  \label{sleq}
\end{equation}%
with the potential 
\begin{equation}
V\left( z\right) =\left( \frac{15}{4}+m_{5}^{2}R^{2}\right) \frac{1}{z^{2}}%
+\lambda ^{2}\left( \lambda ^{2}z^{2}+2\right) .  \label{vsw}
\end{equation}%
This formulation is convenient for evaluating the discrete meson mass
spectrum $m_{n}^{2}=q_{n}^{2}$ associated with the normalizable scalar bulk
modes $\phi \left( m_{n},z\right) $.

At this point, a few comments are in order on the application of current
AdS/QCD duals to three-color (i.e. $N_{c}=3$) physics in general and to
exotics in particular. The semiclassical treatment of the underlying\ string
(or, at weak curvature, gravity) dynamics is justified for large $N_{c}$.
Since calculating even the leading $1/N_{c}$ corrections is beyond present
capabilities, the extrapolation to $N_{c}=3$ has to remain naive and
speculative. This holds in particular for bottom-up approaches where the $%
N_{c}$ dependence enters by comparison of calculated amplitudes with
gauge-theory results \cite{erl05}, for example in bulk coupling and mass
parameters (one finds e.g. $g_{5}^{2}=12\pi ^{2}/N_{c}$ for the gauge
coupling in Eq. (\ref{act})), in the normalization of correlators, or when
encoding the Yang-Mills $\beta $ function. From a formal perspective, the
resulting extrapolation to smaller $N_{c}$ should therefore be regarded as a
working hypothesis. In any case, it provides a consistent $N_{c}$ counting
for the so far calculated observables and a semi-quantitative description of
classical hadron properties (at the 10--30\% level).

One should keep in mind, however, that this naive treatment faces more
challenging tasks in the exotic multiquark\ sector where the $N_{c}$
dependence can be particularly strong and where the extrapolation to $%
N_{c}=3 $ becomes a practical necessity. Indeed, exotics are dissolving in
the $N_{c}\rightarrow \infty $ limit \cite{man96} (tetraquarks perhaps
already for $N_{c}>30-50,$ as unitarized chiral perturbation theory suggests 
\cite{pel04}) and the higher-lying tetraquarks are expected to remain highly
unstable under strong decay (with widths comparable to their masses) even at 
$N_{c}=3$ \cite{jaf77}. Our main focus will be on the lightest scalar
tetraquark nonet, however, which should be relatively narrow (due to the
paucity of open decay channels, limited phase space and partial dynamical
suppression) and hopefully reasonably amenable to the naive $%
N_{c}\rightarrow 3$ extrapolation.

\section{Interpolator and anomalous dimension}

The AdS/CFT dictionary \cite{revs1} relates the normalizable bulk modes $%
\phi $ to those interpolating fields of the dual gauge theory which have
enhanced overlap with the corresponding hadrons. More specifically, in the
scalar sector the conformal dimension $\Delta $ of a meson interpolator
prescribes the mass term in the potential (\ref{vsw}) as

\begin{equation}
m_{5}^{2}R^{2}=\Delta \left( \Delta -4\right)  \label{m}
\end{equation}%
and thereby sets the $z\rightarrow 0$ boundary condition for its dual mode,
i.e. for the solution of Eq. (\ref{sleq}). In the case of ordinary scalar\
mesons the quark-antiquark interpolator%
\begin{equation}
J_{\bar{q}q}^{A}=\bar{q}^{a}t^{A}q^{a}  \label{qqi}
\end{equation}%
is (up to factors)\ the unique choice with $\Delta _{\bar{q}q}=3$. For light
scalar mesons with a dominant tetraquark component, on the other hand, the
larger number of possible couplings between the four quarks \cite{san07}
allows for several potential tetraquark interpolators with the minimal
classical scaling dimension $\Delta _{\bar{q}^{2}q^{2}}=6$. In bottom-up
duals it is therefore not \textit{a priori} clear how to assign states to
these interpolators. In the light of our above comments a natural choice,
containing a good diquark and a good antidiquark, would be%
\begin{equation}
J_{\bar{q}^{2}q^{2}}^{A}=\varepsilon ^{abc}\varepsilon ^{ade}\bar{q}%
^{b}C\Gamma ^{A}\bar{q}^{c}q^{d}C\Gamma ^{A}q^{e}  \label{ti}
\end{equation}%
where $C$ is the charge conjugation matrix, no sum over $A$ is implied and
the spin-flavor matrices $\Gamma $ for the different members of the nonet
may be found e.g. in Refs. \cite{bri05,sch03}. The explicit form (\ref{ti})
is just an illustrative example, however, and will not affect our results.
Indeed, for our purpose it suffices to specify the tetraquark interpolator's
quantum numbers, its scaling dimension and the defining property of maximal
overlap with the tetraquark ground state.

The soft-wall meson spectrum resulting from the $\bar{q}q$ interpolator (\ref%
{qqi}) is \cite{veg08} 
\begin{equation}
m_{\bar{q}q,n}^{2}=4\left( n+\frac{3}{2}\right) \lambda ^{2}  \label{m2qbarq}
\end{equation}%
for $n\geq 0$. The eigenmodes are $\phi _{\bar{q}q}\left( m_{n},z\right)
\propto \left( \lambda z\right) ^{3}\left( z/R\right) ^{-3/2}e^{-\lambda
^{2}z^{2}/2}L_{n}^{\left( 1\right) }\left( \lambda ^{2}z^{2}\right) $ where $%
L_{n}^{\left( \alpha \right) }$ is a generalized Laguerre polynomial \cite%
{abr72}. The naive tetraquark spectrum corresponding to an interpolator like
Eq. (\ref{ti}) with classical scaling dimension $\Delta _{\bar{q}%
^{2}q^{2}}=6 $ (and no anomalous dimension), on the other hand, is 
\begin{equation}
m_{\Delta =6,n}^{2}=4\left( n+3\right) \lambda ^{2}.  \label{m2tetnaive}
\end{equation}%
The radial excitations are thus likewise organized into a linear square-mass
trajectory with the universal slope $4\lambda ^{2}$. The corresponding
normalizable eigenmode solutions are now 
\begin{equation}
\phi _{\Delta =6}\left( m_{n},z\right) =N_{\Delta =6,n}\left( \lambda
z\right) ^{6}\left( \frac{z}{R}\right) ^{-3/2}e^{-\lambda
^{2}z^{2}/2}L_{n}^{\left( 4\right) }\left( \lambda ^{2}z^{2}\right) .
\end{equation}%
The intercept of the $\Delta =6$ trajectory is twice as large as that of its 
$\Delta =3$ counterpart, however, which implies $m_{\Delta =6,0}^{2}/m_{q%
\bar{q},0}^{2}=2$ for the ground states. This exemplifies the general case:
in the absence of bulk interactions and when ignoring anomalous dimensions
(which is so far common practice in AdS/QCD applications), the\ holographic
meson-mass predictions increase with the interpolator dimension $\Delta $
(as $m_{n}^{2}=4\left( n+\Delta /2\right) \lambda ^{2}$ in the dilaton
soft-wall for $d=4$).

The above pattern adequately reproduces the empirical mass hierarchy among
the less strongly bound $\bar{q}q$ mesons (with $\Delta =3$) and scalar
glueballs (with $\Delta =4$). The exceptionally strong attraction required
to reduce the masses of the lightest tetraquark nonet below those of the
scalar $\bar{q}q$ ground state, on the other hand, is obviously not
reflected in this minimal description. In the following we will examine
whether the neglect of the tetraquark interpolator's anomalous dimension may
be responsible for this shortcoming. As already mentioned, this would not be
unexpected if the additional binding originates from the maximally
attractive diquark channel because diquark-content-dependent anomalous
dimensions of the nucleon interpolators were found to describe good-diquark
binding effects in the light baryon spectrum \cite{for09} (on the basis of
the \textquotedblleft metric soft-wall\textquotedblright\ gravity dual of
Ref. \cite{for07}). Indeed, the anomalous dimension of the tetraquark
interpolator will generally depend on its diquark content as well, and thus
on the diquark content of the scalar meson with which it has enhanced
overlap. Moreover, if two quarks form a good diquark whenever possible, and
if good diquarks systematically reduce the masses of the hadrons which they
dominate (compared to those containing other types of diquarks), one would
expect this behavior to have a universal representation in the gravity dual.

Holographically, the inclusion of the anomalous dimension $\gamma $ turns
the mass term (\ref{m}) for the bulk modes dual to the $\Delta _{\bar{q}%
^{2}q^{2}}=6$ interpolator into%
\begin{equation}
m_{5}^{2}\left( z\right) R^{2}=\left[ 6+\gamma \left( z\right) \right] \left[
2+\gamma \left( z\right) \right] .  \label{m2}
\end{equation}%
(In more complex gravity duals, mass corrections may further arise from
couplings to additional, e.g. condensate-related bulk fields.) The $z$
dependence of $\gamma $ is inherited from its renormalization group (RG)
flow $\gamma \left( \mu \right) $ since the AdS/CFT dictionary translates
the RG scale $\mu $ into the inverse of the fifth dimension, i.e. $\mu \sim
1/z$ \footnote{%
We\hspace{-0.2mm} shall\hspace{-0.2mm} tentatively\hspace{-0.2mm} assume%
\hspace{-0.2mm} that\hspace{-0.2mm} the RG mixing\hspace{-0.2mm} among
anomalous\hspace{-0.2mm} dimensions remains\hspace{-0.2mm} small\hspace{%
-0.2mm}\hspace{-0.2mm} in\hspace{-0.2mm} the \hspace{-0.2mm}restricted 
\hspace{-0.2mm}$z$ (resp.\hspace{-0.2mm} $\mu $)\hspace{-0.2mm} range\hspace{%
-0.2mm} $z\lesssim 1/\lambda $ of\hspace{-0.2mm} interest\hspace{-0.2mm}%
\hspace{-0.2mm}.}. Ideally, $\gamma \left( z\right) $ should be obtained
from (both perturbative and nonperturbative) QCD and, as any other
information from the gauge-theory side, implemented into the gravity dual in
bottom-up fashion. However, at present this approach is ruled out by the
absence of QCD information on $\gamma \left( z\right) $. As a preliminary
substitute, we will therefore resort to a reasonable ansatz for $\gamma
\left( z\right) $ and restrict it as tightly as possible by consistency,
stability and physics requirements. This strategy cannot directly relate the
ensuing mass corrections (\ref{m2}) to the anomalous dimension in QCD, but
it will provide quantitative orientation and general insights, especially on
how much tetraquark binding the anomalous-dimension mechanism can generate
in principle, and even lead to a rather unique semi-quantative guess for $%
\gamma $ which is essentially independent of the specific AdS/QCD dynamics (%
\ref{act}) adopted to derive it. Nevertheless, this estimate will have to be
checked against -- and in case of significant disagreements replaced by --
QCD results for $\gamma $ (e.g. from the lattice) when they eventually
become available.

A glance at the soft-wall potential (\ref{vsw}) with the generalized mass
term (\ref{m2}) for $\Delta _{\bar{q}^{2}q^{2}}=6$ reveals that even a
scale-independent anomalous dimension $\gamma <-3$ would reduce its
repulsion sufficiently to bring the tetraquark ground-state mass down below
that of the $\bar{q}q$ ground state. However, it would do the same for all
radial excitations as well, whereas phenomenologically higher-lying
tetraquark excitations (with the exception of the first one if interpreted
as a recurrence of the tetraquark ground state \cite{mai07}) should be
pushed up toward or beyond the corresponding $\bar{q}q$ excitations, to be
broad enough to explain the absence of additional scalar nonets in the
measured meson spectrum. This suggests to check whether a suitably running $%
\gamma \left( z\right) $ can accommodate the phenomenological situation. As
it turns out, the necessary qualitative\ requirements may be cast into the
simple power-law behavior%
\begin{equation}
\gamma \left( z\right) =-az^{\eta }+bz^{\kappa }  \label{gam}
\end{equation}%
with $\kappa \gg \eta $, so that the first (second) term modifies mainly the
small-$z$ (large-$z$) region of the Sturm-Liouville potential. Moreover, the
coefficients and exponents in Eq. (\ref{gam}) can be almost quantitatively\
determined by the underlying physics. To show this, we split the soft-wall
potential (\ref{vsw}) with the generalized mass term (\ref{m2}) as 
\begin{equation}
V_{q^{2}\bar{q}^{2}}\left( z\right) =\left( \frac{15}{4}+12\right) \frac{1}{%
z^{2}}+\lambda ^{2}\left( \lambda ^{2}z^{2}+2\right) +\Delta V\left(
z\right)   \label{v4q}
\end{equation}%
into the uncorrected (i.e. $\gamma =0$) potential with the spectrum (\ref%
{m2tetnaive}) and the finite-$\gamma $ correction 
\begin{equation}
\Delta V\left( z\right) =\gamma \left( z\right) \left[ \gamma \left(
z\right) +8\right] \frac{1}{z^{2}}.  \label{DelV}
\end{equation}%
Maintaining asymptotically linear trajectories and enough repulsion to move
the masses of the higher excitations upward restricts the exponent $\kappa $
to $\kappa \lesssim 2$. Approximate slope universality \cite{bug04} further
demands $\left\vert b\right\vert \ll 1$. The collapse of solutions to the
mode equation (\ref{sleq}) into the AdS$_{5}$ boundary at $z\rightarrow 0$ 
\cite{lan91} is avoided for $\eta >0$ (see below), and maximal attraction at
smaller $z$, where the $z^{-2}$ enhancement reduces the ground-state mass
most effectively, demands $\eta \ll 1$. As a consequence, Eq. (\ref{gam})
does not affect the AdS/CFT boundary condition of the tetraquark bulk modes
for $z\rightarrow 0$ since $\gamma \left( 0\right) =0$, in accord with
asymptotic freedom. Maximizing the overall attraction of the potential
correction (\ref{DelV}), finally, requires $a\simeq 4$ (see below). It is
worth noting that anomalous dimensions of a $z$ dependence qualitatively
similar to Eq. (\ref{gam}) (in the region of interest) have recently been
found in dual\ backgrounds which are of holographic RG-flow type \cite{mue10}%
. (For a holographic estimate of the RG flow of an effective QCD coupling
see Ref. \cite{bro10}.)

To put the above arguments in favor of an unusually large anomalous
dimension for the tetraquark interpolator into perspective, we recall that
in AdS/QCD applications to \emph{ordinary} hadrons anomalous dimensions are
generally neglected. This provides a decent approximation as long as the
anomalous dimensions of the corresponding two- and three-quark interpolators
are much smaller than their classical scaling dimensions. In the UV, the
latter is borne out by perturbative QCD results which show a weak,
logarithmic RG scaling behavior. In the IR, on the other hand, where much
less QCD information is currently available, it seems that anomalous
dimensions can become relevant in some channels and generate notable
holographic effects, as studied in Ref. \cite{for09}. Nevertheless, as a
rule one does not expect them to reach magnitudes comparable to the
classical scaling dimensions. (Otherwise results of standard AdS/QCD would
have to be revised). Our above arguments suggest that the exotic tetraquark
channel provides an exception to this rule in which extraordinarily strong
dynamical effects (related to good-diquark binding) render the anomalous
dimension non-negligible even to leading approximation.

\section{Results and discussion \label{res}}

Several useful and rather generic features of the tetraquark mass spectrum
in potentials of the type (\ref{v4q}), (\ref{DelV}) can be understood
without actually solving the eigenvalue problem (\ref{sleq}). First, any $%
\gamma \left( z\right) $ which renders the correction (\ref{DelV}) negative
and decreases the power of $z$ below $-2$ for $z\rightarrow 0$ would cause a
physically unacceptable \textquotedblleft collapse to the
center\textquotedblright\ instability of the Sturm-Liouville solutions \cite%
{lan91}. In our context, the \textquotedblleft center\textquotedblright\ at $%
z=0$ corresponds to the AdS$_{5}$ boundary, and this collapse is avoided by
keeping $\eta >0$. An instability may appear even if the singularity remains 
$\propto 1/z^{2}$, however. Indeed, as long as the coefficient of this
singularity is negative and exceeds a critical magnitude, the negative
potential energy will overcome the stabilizing kinetic (or localization)
energy. A straightforward analysis of the small-$z$ behavior of the
solutions of Eq. (\ref{sleq}) in the potential (\ref{v4q}) for general
boundary dimension $d$ shows that such a collapse is prevented as long as 
\begin{equation}
\frac{d^{2}}{4}+\tilde{\Delta}\left( \tilde{\Delta}-d\right) >0,  \label{bfb}
\end{equation}%
where we abbreviated $\tilde{\Delta}\equiv \Delta +\gamma $. In fact, taking 
$\tilde{\Delta}$ constant turns the inequality (\ref{bfb}) into the
Breitenlohner-Freedman (BF) bound $m_{5}^{2}R^{2}>-d^{2}/4$ \cite{bre82}. A
violation of\ this bound thus causes the ground-state wavefunction to
disintegrate and its\ eigenvalue to become $m_{0}^{2}=-\infty $. This is a
holographic signal for a vacuum instability in the boundary gauge theory.
Our form (\ref{DelV}) of $\Delta V$, however, which arises from the
anomalous-dimension correction (\ref{m2}), has the lower bound%
\begin{equation}
\Delta V\left( z\right) \geq -\frac{16}{z^{2}}  \label{DV}
\end{equation}%
(for $d=4$ and all $z$). Hence it preserves the BF bound and prevents a
collapse from the outset. The inequality (\ref{DV}) is saturated for
(constant) $\tilde{\Delta}=2$\ or $\gamma =-4$. This generates the lower
bound%
\begin{equation}
m_{\bar{q}^{2}q^{2},0}\geq m_{\Delta =2,0}=2\lambda  \label{mbd}
\end{equation}%
for the ground-state mass. Therefore $2\lambda $ is the minimal tetraquark
mass which can arise from the bulk mass term (\ref{m2}) with any anomalous
dimension. (A coupling of the tetraquark's dual mode to an additional bulk
field may release the bound (\ref{DV}) for $z>0$ and thus generate lower
masses.)

\begin{figure}[tbp]
\begin{center}
\includegraphics[height = 9cm]{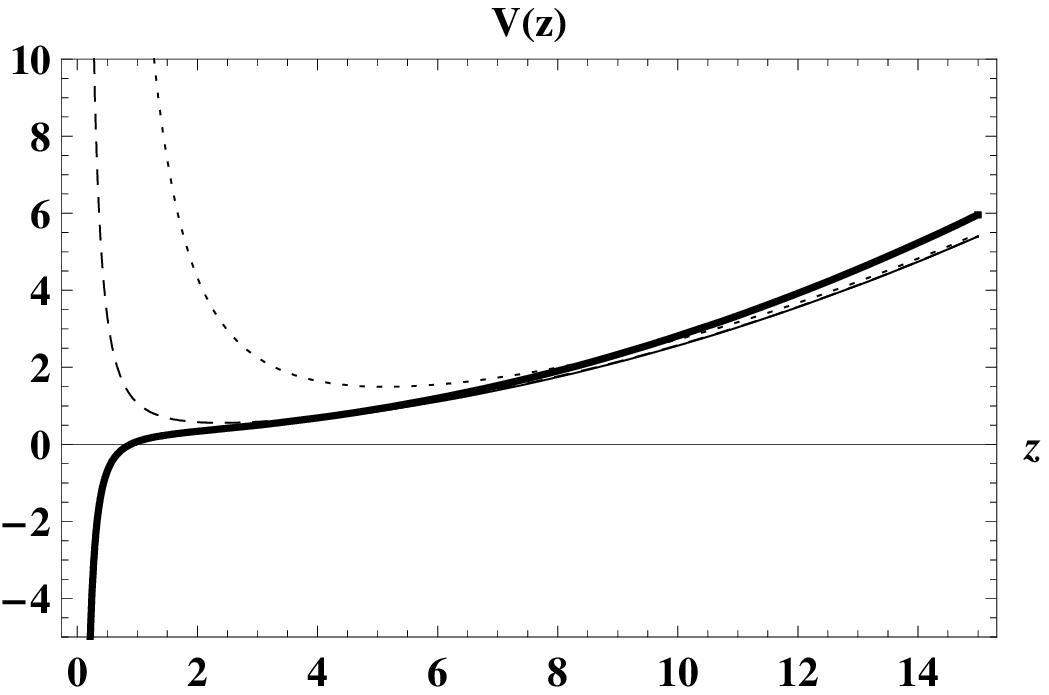}
\end{center}
\caption{The Sturm-Liouville potential (\protect\ref{v4q}) of the
tetraquark's dual mode for the anomalous dimension (\protect\ref{gam}) with $%
a=4$, $b=0.05$, $\protect\eta =0.001$ and $\protect\kappa =2$ (thick line).
For comparison, we also plot the potential for the four-quark interpolator
with $\Delta =6,$ $\protect\gamma =0$ (dotted line) and for the $\bar{q}q$
interpolator with $\Delta =3$ (dashed line). In addition, we plot the
potential for $\Delta =2$ (thin line) which saturates the
Breitenlohner-Freedman bound. }
\label{fig1}
\end{figure}

As a result, we have shown that a suitable ansatz for the tetraquark
interpolator's anomalous dimension can indeed generate an extraordinarily
large holographic binding energy which drives the lightest tetraquark mass
from about 40\% above down to around 20\% \emph{below} the mass $m_{q\bar{q}%
,0}=\sqrt{6}\lambda $ (cf. Eq. (\ref{m2qbarq})) of the $\bar{q}q$ ground
state. To provide a concrete example for a $\gamma \left( z\right) $ of the
form (\ref{gam}) which results in a bulk-mode potential with nearly maximal
attraction for the ground state, we choose the coefficients inside the
narrow range established by the criteria given above as $a=4$, $b=0.05$, $%
\eta =0.001$ and $\kappa =2$. The resulting potential (\ref{v4q}) with a
typical soft-wall IR scale $\lambda =0.425$ GeV is plotted in Fig. \ref{fig1}%
. (For comparison, we also show the potentials corresponding to the $\Delta
=3$ and $\Delta =6$ interpolators without anomalous dimensions, as well as
to the one with $\Delta =2$ which saturates the BF bound.) Numerical
integration of Eq. (\ref{sleq}) in the plotted potential indeed yields a
ground-state mass close to the bound (\ref{mbd}): $m_{\bar{q}%
^{2}q^{2},0}=2.11\lambda $. Although the particular ansatz (\ref{gam}) for $%
\gamma \left( z\right) $ is not unique, Fig. \ref{fig1} reveals that the
bounds deduced above determine the behavior of the (almost) maximally
attractive potential in a practically model-independent manner.

\begin{figure}[tbp]
\begin{center}
\includegraphics[height = 9cm]{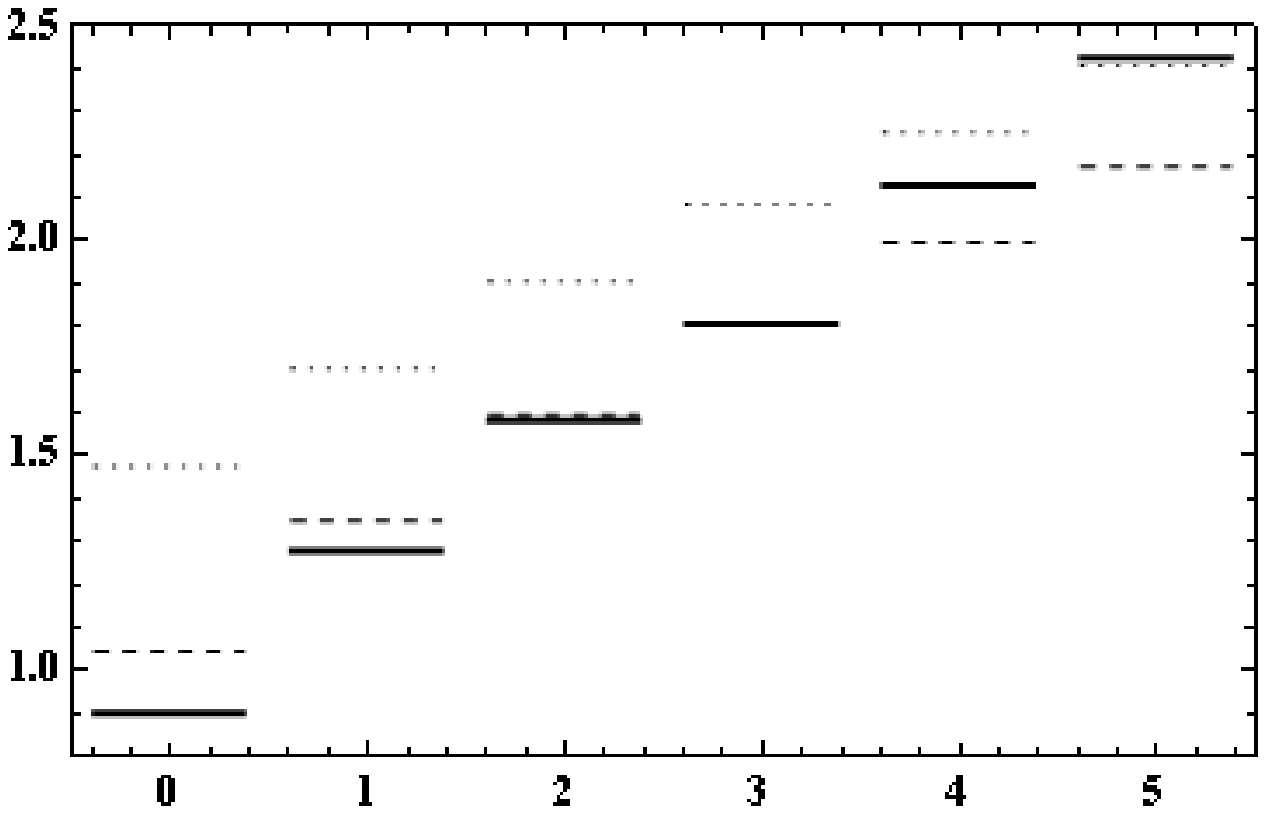}
\end{center}
\caption{The ground state ($n=0$) and the first five excitations of the
tetraquark mass spectrum in the potential of Fig. \protect\ref{fig1} (thick
bars). We further plot the spectra for the $\bar{q}q$ interpolator (Eq. (%
\protect\ref{m2qbarq}), dashed bars) and for the $\Delta =6$ interpolator
(Eq. (\protect\ref{m2tetnaive}), dotted bars) with vanishing anomalous
dimensions.}
\label{figspec}
\end{figure}

We now turn to the radial excitation spectrum of the tetraquark. The masses
of the first five excited states (in the potential of Fig. \ref{fig1}) are
shown in Fig. \ref{figspec}. In this part of the spectrum, the square masses
increase considerably faster than their asymptotic, linear trajectory 
\begin{equation}
m_{n}^{2}\overset{n\rightarrow \infty }{\longrightarrow }4n\sqrt{\lambda
^{4}+b^{2}}\overset{\left\vert b\right\vert \ll \lambda ^{2}}{\simeq }%
4\lambda ^{2}n
\end{equation}%
with almost universal slope $4\lambda ^{2}$. In particular, they start to
surpass the corresponding masses on the $\bar{q}q$ trajectory around the
second excitation level (and those on the uncorrected $\Delta =6$ trajectory
from about the fifth excitation onward). As intended, the higher-lying
tetraquarks can  therefore become broad enough to escape detection. Whereas
a fully quantitative comparison of Fig. \ref{figspec} with the experimental
scalar mass spectrum as well as a detailed analysis of the numerically
generated eigenmodes and decay constants would seem premature in our
explorative framework, a few additional and probably rather robust features
of the resulting spectrum are noteworthy. Interpreting the second scalar
nonet around 1.5 GeV as predominantly the lightest $\bar{q}q$ state would
imply the phenomenological mass ratio $m_{\bar{q}^{2}q^{2},0}/m_{\bar{q}%
q,0}\sim 0.8/1.5\simeq 0.53$ which is somewhat larger than what the
tetraquark interpolator's anomalous dimension by itself can generate. (The
alternative interpretation of the second nonet as a tetraquark recurrence 
\cite{mai07}, on the other hand, would raise the question of how to assign
the lightest $\bar{q}q$ state.) The remaining part of the ground-state mass
ratio may e.g. be supplied by the anomalous dimension of the $\bar{q}q$
interpolator or by couplings of the dual tetraquark modes to additional
background fields. An adequate description of molecular-binding or
meson-cloud effects could require to take bulk-mode interactions into
account, moreover, and the resonance widths will further modify the masses.

A reassuring feature of our anomalous-dimension-mediated tetraquark binding
mechanism is that it will work in essentially the same fashion in any of the
current AdS/QCD gravity duals. Indeed, $\gamma $ enters the bulk dynamics
and hence the tetraquark mode potential only through the mass term of Eq. (%
\ref{vsw}) which universally appears in all AdS/QCD duals \footnote{%
This is because the geometry of AdS/QCD duals approaches exact AdS$_{5}$ for 
$z\rightarrow 0$ in order to ensure conformality of the dual gauge theory in
the UV.} and which\ is model-independently fixed by the AdS/CFT dictionary
in Eq. (\ref{m2}). As a consequence, the anomalous-dimension-induced
correction (\ref{DelV}) and its impact on the tetraquark spectrum are
independent of the specific AdS/QCD dynamics under consideration. The $%
\gamma \left( z\right) $ determined above should therefore generate a light
scalar tetraquark with substantially more massive radial excitations in any
AdS/QCD dual. Nevertheless, it would still be interesting to explore the
quantitative impact of the correction (\ref{DelV}) in other and especially
backreacted versions of AdS/QCD (such as Ref. \cite{dep209}).

The holographic picture of light tetraquarks emerging from our above results
can be extended and tested further, e.g. by implementing flavor-symmetry
breaking and bulk-mode interactions, analyzing the ensuing decay channels
and studying pertinent correlators (which may contain both $\bar{q}q$ and $%
\bar{q}^{2}q^{2}$ interpolators). It would also be interesting to
investigate the behavior of exotic multiquarks in a topological expansion
with large numbers of both colors ($N_{c}$) and flavors ($N_{f}$) where $%
N_{f}/N_{c}$ is kept fixed \cite{ven76}.

\section{Summary and conclusions}

We have discussed several foundational aspects of light scalar tetraquark
physics from a holographic perspective. In particular, we have identified an
AdS/QCD representation for the exceptionally large tetraquark binding energy
in terms of resolution-dependent bulk-mass corrections for the corresponding
dual modes. These corrections are argued to originate (at least in part)
from the running anomalous dimension of the four-quark interpolator which
encodes the diquark content of the tetraquark. As a consequence, they are
independent of the specific AdS/QCD dynamics under consideration. The
minimal achievable tetraquark mass is determined by the
Breitenlohner-Freedman bound, furthermore, which prevents the tetraquark's
dual mode from collapsing into the anti-de Sitter boundary.

In order to provide a quantitative example for this binding mechanism, we
have implemented it into the dilaton-soft-wall gravity dual for holographic
QCD. The bulk mass corrections which generate nearly maximal attraction in
the tetraquark channel are found to be largely determined by consistency and
stability requirements, and the minimal tetraquark mass is fixed at twice
the dilaton-induced infrared scale. As a result, the generated attraction is
indeed strong enough to render the tetraquark the lightest scalar meson,
about 20\% lighter than the $\bar{q}q$ ground state. To meet quantitative
phenomenological expectations would probably require some additional
contribution, however, which may originate e.g. from the anomalous dimension
of the $\bar{q}q$ interpolator or from additional bulk fields and
interactions. The higher tetraquark excitations turn out to be heavier than
their $\bar{q}q$ counterparts, finally, so that their increasing widths (due
to increasingly many open decay channels, a larger phase space and potential
dynamical enhancements) should prevent them from generating more low-lying
scalar resonances than experimentally seen.

We would like to thank Dietmar Ebert and Eberhard Klempt for their questions
regarding a holographic description of light scalar tetraquarks. Financial
support from the Deutsche Forschungsgemeinschaft (DFG) is also acknowledged.

\newpage

\end{document}